\documentclass[final,authoryear,5p,times,twocolumn]{elsarticle}

\usepackage{amssymb}
\usepackage{epsfig}
 \usepackage{subfigure}
\usepackage{graphicx}
\usepackage{rotating}
\def\astrobj#1{#1}

\journal{New Astronomy}

\begin{document}

\begin{frontmatter}



\title{Searching for short-period variable stars in the direction of Coma Berenices and Upgren~1 open clusters: 
Melotte~111~AV~1224 a new eclipsing binary star}


\author[oan]{L. Fox Machado\corref{cor}}
\ead{lfox@astrosen.unam.mx}

\author[oan]{R. Michel}

\author[oan]{M. Alvarez}

\author[ia]{J.H. Pe\~na}

\cortext[cor]{Corresponding author. Tel.: +52 6461744580; fax +52
6461744607 }

\address[oan]{Observatorio Astron\'omico Nacional, Instituto de
Astronom\'{\i}a -- Universidad Nacional Aut\'onoma de M\'exico, Ap.
P. 877, Ensenada, BC 22860, M\'exico}

\address[ia]{Instituto de Astronom\'{\i}a  -- Universidad Nacional Aut\'onoma de M\'exico,
Ap. P. 70-264, M\'exico, D.F. 04510, M\'exico}

\begin{abstract}
 We report the results of CCD  photometric observations in the direction of the Coma Berenices and 
Upgren 1 open clusters with the aim at searching for new short-period variable stars.  
A total of 35 stars were checked for variability. 
As a result of this search the star designated in the SIMBAD database as  
 Melotte~111~AV~1224 was found to be a new eclipsing binary star. Follow-up Str\"omgren photometric
and spectroscopic observations allowed us to derive the spectral type, distance, reddening
and  effective temperature of the star.  A preliminary analysis of the binary light curve  
was performed and the parameters of the orbital system were derived. 
From the derived physical parameters we conclude that Melotte~111~AV~1224 is most likely a W-UMa eclipsing binary
that is not a member of the Coma Berenices open cluster.
On other hand, we did not find evidence of
brightness variations in the stars NSV~5612 and NSV~5615 previously catalogued as
variable stars in Coma Berenices open cluster.

\end{abstract}


\end{frontmatter}


\section{Introduction}
\label{sec:int}

The study of short period variable stars in open clusters
 is fundamental in stellar evolution. 
Since all the cluster members are assumed to have the same interstellar
reddening, distance, age and chemical abundance, it is possible to
put strong constraints on these physical parameters of the
cluster variables in asteroseismic model calculations  [e.g. \citet{fox1,fox2}].

Besides the great success of multichannel photoelectric photometers to study short
period variables in open clusters  [e.g. \citealt{costa}; \citet{fox3,fox4}; \citet{li,li1}], 
CCD technique working in the time-series photometry mode 
has been preferred for open clusters observations (e.g. \citealt{kang}; 
\citealt{kopacki}; \citealt{luo}; \citealt{choo}).  Indeed a CCD camera allows to obtain high precision photometric data
by measuring the program star and 
reference stars simultaneously 
in the CCD's field of view (FOV hereafter)
under the same weather and instrumental conditions.
The number of stars observed simultaneously with a CCD camera may vary from a couple, 
in case of a small FOV and sparse fields (e.g \citealt{baran}), up to thousands for mosaic CCDs 
pointing near the plane of the Milky Way. 

Taking the advantage of CCD cameras we have carried out a search for new short period variable stars in the direction
of the Coma Berenices and Upgren 1 open clusters. Coma Berenices (Melotte~111 hereafter, 
RA\,$=12^{\rm h}$23$^{\rm m}$, DEC\,$=+26^{\rm o}$00$^{\prime}$, J2000.0 )
 is the second closest open cluster to the Sun being more distant than the Hyades 
($\sim$ 45 pc) but closer than the Preasepe ($\sim$ 180 pc).
The  \textit{Hipparcos} intermediate astrometry data place it at 
  $d=86.7 \pm 0.9$~pc \citep{vanleeuwen}. 
  An  almost solar metallicity has been derived in the cluster (e.g. [Fe/H] $=-0.065 \pm 0.021$~dex, \citealt{cayrel},
[Fe/H] $=0.052 \pm 0.047$~dex, \citealt{friel}).
 The age of the cluster is 
estimated at between 400 and 600 Myr (e.g. \citealt{bounatiro}). 
Several investigations which have addressed the 
stellar population in Melotte~111  support the fact that the cluster has relatively few members and, particularly that it is poor in low-mass stars.
(e.g., \citealt{artyukhina}; \citealt{argue}; \citealt{deluca}; \citealt{bounatiro}). 
Several searches for new low-mass cluster members 
have been recently performed but without more success than early studies (e.g. \citealt{casewell}; \citealt{mermilliod}; \citealt{melnikov};
\citealt{terrien}). 
In particular \cite{terrien} have addressed the membership of the stars in direction of Coma Berenices
using SDSS III APOGEE radial velocity measurements, confirming just eight K/M dwarf new candidate members of the cluster.
Given that Melotte~111 is relatively sparse and the  
verification of membership in the cluster has been challenging, the detection of new variable stars 
 that are members of the cluster is very important. 
Meolotte~111 covers about 100 ${\rm deg}^{2}$ on the sky, but its central part occupies only about 25 ${\rm deg}^{2}$. 
The core of the cluster is estimated between 5-6 pc \citep{odenkirchen}.
The catalogue of variable stars in open cluster by \cite{zejda} lists
57 variable stars belonging to the Melotte~111 open cluster. 

Upgren 1 (RA\,$=12^{\rm h}$35$^{\rm m}$, DEC\,$=+36^{\rm o}$22$^{\prime}$, J2000.0) is an 
association of seven F-type stars located
at a distance of $\sim$ 117 pc and considered to be
 a remnant of an old galactic open cluster (\citealt{upgren}, \citealt{upgren1}).
 
 \cite{gatewood}  studied  these stars with 
a multichannel astrometric photometer 
and proposed that the cluster is composed
  of two dynamically different groups. We have observed four stars in this association. 

The paper is organized as follows. In Sect.~\ref{sec:obs}, the acquisition of the data
and the description of the observations are presented.
In Sect.~\ref{sec:lcur}, the analysis of differential light curves and the fields observed
are discussed. Sect.~\ref{sec:av1224} is devoted
to the analysis of the light curve  and the physical parameters of  Melotte~111~AV~1224.
 In Sect.~\ref{sec:con} we summarize our conclusions.

\section{Observations and data reduction}
\label{sec:obs}

\begin{figure}[!t]
\includegraphics[width=6.0cm,height=6.0cm]{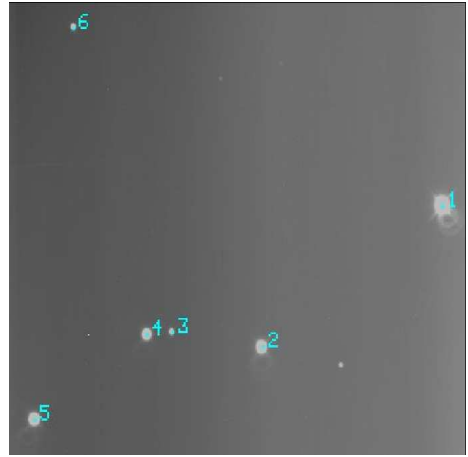}  
\caption{CCD FOV~1. The following targets were observed:
1-NSV~5613 (BD+27~2129) , 2-NSV~5615 (BD+27~2130), 3-USNO B1 1167-0214141,
4-Melotte~111~AV~1616, 5-Melotte~111~124, 6-Melotte~111~AV~1625. North is up and East is left. } \label{fig:field1}
\end{figure}

\begin{figure}[!t]
\includegraphics[width=6.0cm,height=6.0cm]{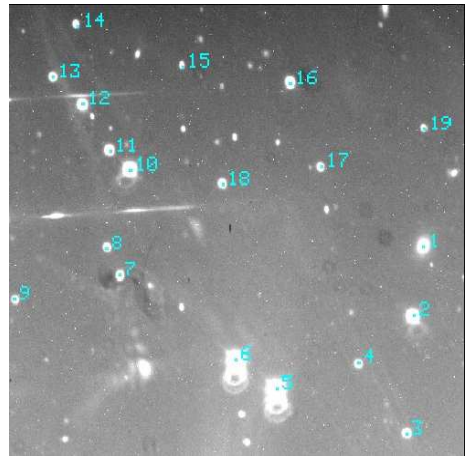}  
\caption{CCD FOV 2. The following targets are identified:
 1-IC~3194 (Galaxy), 2-Melotte~111~AV~1176, 3-USNO B1 1151-0193085,
4-USNO B1 1150-0195002, 5-Melotte~111~AV~1192,  6-Melotte~111~AV~1196,
7-USNO B1 1151-0193137, 8-USNO-B1 1151-0193141, 9-SSDS J122138.25+250714.5,
10-Melotte~111~AV~1204, 11-Melotte~111~AV~1207, 12-SSDS J122135.63+251028.0,
13-SDSS J122135.43+251039.0, 14-SDSS J122133.76+251127.7, 
15-SDSS J122133.76+251127.7, 16-SDSS J122121.31+251048.2, 17-SDSS J122123.35+250859.9,
18-SDSS J122123.35+250859.9, 19-SDSS J122108.89+250949.1. North is up and East is left.
} \label{fig:field2}
\end{figure}

\begin{figure}[!t]
\includegraphics[width=6.0cm,height=6.0cm]{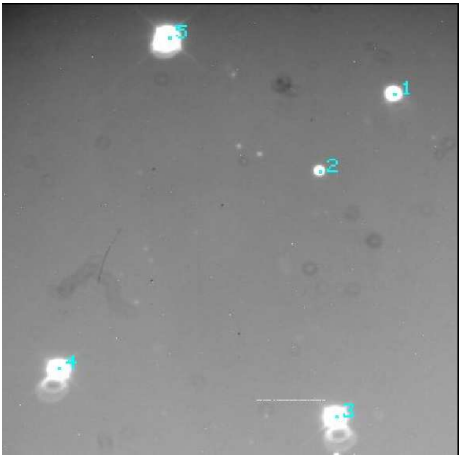}  
\caption{CCD FOV 3. The following targets were observed:
1-BD+25~2296,2-GSC 02530 02027,  3-HD~109509, 4-HD~109542,, 5-HD~109530. North is up and East is left.
} \label{fig:field3}
\end{figure}

\begin{figure}[!t]
\includegraphics[width=6.0cm,height=6.0cm]{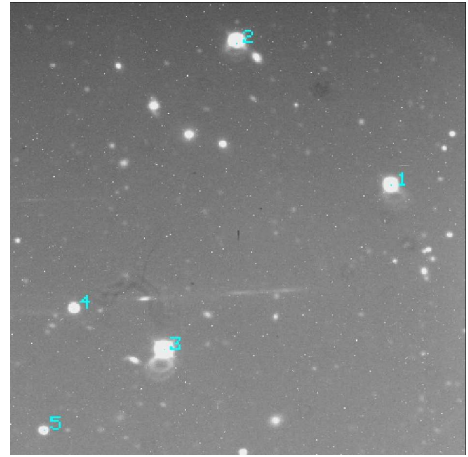}  
\caption{CCD FOV 4. The following targets are listed:
1-Melotte~111~AV~1224, 2-Melotte~AV~1236, 3-Melotte~111~AV~1248,
4-SDSS J122202.82+245901.5, 5-USNO-B1 1149-0192498. North is up and East is left.
 } \label{fig:field4}
\end{figure}

The CCD  observations have been made with the 0.84-m
f/15 Ritchey-Chr\'etien telescope at Observatorio Astron\'omico Nacional at Sierra San Pedro M\'artir (OAN-SPM),  
during  ten consecutive nights, between April 11 and 20, 2009. The 
telescope hosted the filter-wheel `Mexman' with the Marconi (E2V) CCD camera,
which has a 2048 $\times$ 2048 pixels array, with a pixel size of 15 $\times$ 15 $\mu$m$^{2}$.
The gain and readout noise of the CCD camera are 1.8 e$^{-}$/ADU and 7.0 e$^{-}$,
respectively. The typical field of view in this configuration  
is about 8$^{\prime}$ $\times$ 8$^{\prime}$ arcmin$^{2}$  with the
scale of 0. 28$^{\prime \prime}$/pixel.  
To search for short-period variable stars in direction
of Melotte~111  we observed three fields centered at the following
coordinates  $\alpha_{\rm J2000.0}=12^{\rm h}21^{\rm m} 22^{\rm s}.0$ $\delta_{\rm J2000.0}= +25^{\circ}08^{\prime}31.0^{\prime \prime}$
; $\alpha_{\rm J2000.0}=12^{\rm h}26^{\rm m}05^{\rm s}.0$ $\delta_{\rm J2000.0}=+26^{\circ}46^{\prime}00.0^{\prime \prime}$  and 
 $\alpha_{\rm J2000.0}=12^{\rm h}21^{\rm m}57^{\rm s}.0$, $\delta_{\rm J2000.0}= +25^{\circ}00^{\prime}00.0^{\prime \prime}$.
The images of these FOVs are shown in Figures~\ref{fig:field1},~\ref{fig:field2}, ~\ref{fig:field4} respectively.
The field in direction of Upgren 1 was centered at the  coordinates 
$\alpha_{\rm J2000.0}=12^{\rm h}35^{\rm m}05^{\rm s}.0$, $\delta_{\rm J2000.0}= +36^{\circ}22^{\prime}17.0^{\prime \prime}$ 
(Fig.~\ref{fig:field3}).
 In each figure the target stars are labeled  with numbers and  their corresponding  identifications 
are given in the caption of the figure.  
The log of observations is shown in Table~\ref{tab:log} where the dates, FOV, HJD start, HJD end,
filters, exposure times and number of frames are listed.  We note that the star designation  Melotte~111~AV refers to 
the star running number of the astrometric catalogue for the area of Coma Berenices by \cite{abad}.

Sky flats,
dark and bias exposures were taken each night.
All CCD images were preprocessed to correct overscan, trim unreliable and useless regions, 
subtract bias frames, correct flat fielding and reject cosmic rays using
the IRAF/CCDRED package. Then, instrumental magnitudes of the stars  were computed using the
 point spread function fitting method of the IRAF/DAOPHOT package \citep{massey}. The photometry is gathered in 
tables where frame No., HJD, airmass, UT, photometric instrumental magnitudes and photometric errors
are included. Typical internal errors of the single
frame photometry for stars are of about
0.001 mag in the bands used. The differential light curves  were derived on a star-to-star basis 
computing the difference in magnitude of one star with respect to the others. In that way
each star was checked for variability relative to at least two reference stars.

\section{Analysis of differential light curves}
\label{sec:lcur}

\begin{figure*}[!t]
\begin{center} 
\includegraphics[width=8.0cm,height=10.0cm]{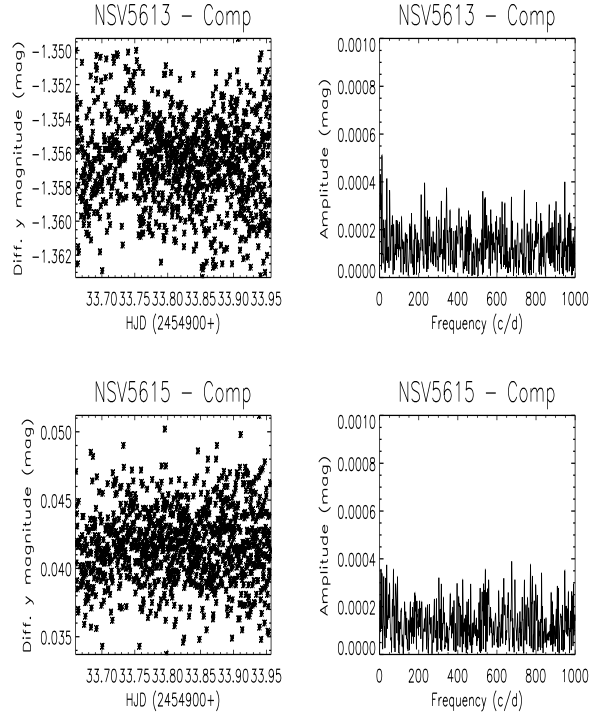} 
\caption{Top: Light curve and amplitude spectrum of NSV~5613.  Bottom:
Light curve and amplitude spectrum of NSV~5615.}
\label{fig:nsv}
\end{center}
\end{figure*}

We proceeded to search for short-period variable stars in each observed field 
using the differential CCD time-series photometry obtained in the previous section.
Our calculations for a V=15~mag star indicated that with 8 hours of data we can achieve a
detection threshold around 1 mmag (millimagnitude) level. Assuming a 4$\sigma$ definition for the 
threshold, the noise level should be 0.25 mmag. This detection threshold is very typical
for most ground-based observations.

The search for stellar pulsations was done in two steps. First, 
all differential light curves were visually inspected for the presence of obvious variability.
In this way we searched the light curves for features like eclipsing binaries, 
planetary transits, flares and high amplitude pulsations. 
Second, all light curves were subject to Fourier analysis. This latter step is very convenient in analysis
of periodicities. It may  uncover a periodic change with either a very tiny amplitude not easily 
seen directly in the light curves or a short period  like those present in pulsating white dwarfs
or sdB type variables.  We calculated Fourier transform up to the Nyquist frequency.

Some comments on the observed field can be made.
The FOV~1  (Fig.~\ref{fig:field1}) includes six stars.
Two stars, NSV~5613 (BD+27~2129) and NSV~5615 (BD+27~2130), 
 have been classified  in the literature
as  variable stars of Melotte~111.  NSV~5613 was reported as suspected variable of Melotte~111 by \cite{golay} while
NSV~5615 has been classified as a RR Lyrae type variable star by \cite{archer}. Since then 
no new observations of these targets has been reported to the best of our knowledge. These stars are also listed in
the International Variable Star Index (VSX)\footnote{http://www.aavso.org/vsx/} database  
of the American Association of Variable Star Observers (AAVSO). Moreover
both stars are included in the catalogue of variable stars in open clusters
  by \cite{zejda}, but no period information is given for them.
We have observed both targets and the adjacent field stars  through a Str\"omgren $y$ filter 
with 30 sec of exposure times which correspond to a Nyquist frequency of 1980 c/d (see Table~\ref{tab:log} for details).
The light curves and amplitude spectra of these two stars are shown in Fig.~\ref{fig:nsv}.
As can be seen there is no evidence of periodic variations either in the light curves
or in the amplitude spectra. 
We conclude that these stars have been wrongly classified as variables, 
as they do not pulsate at all.  
 
 The FOV~2 (Fig.~\ref{fig:field2}) is the most  crowded field we have observed.
 The light curves of the 19 targets numbered in the figure has been derived.
 The following stars are listed as  members of Melotte~111 in the SIMBAD database: Melotte~111~AV~1176,  
Melotte~111~AV~1192,  Melotte~111~AV~1196,  Melotte~111~AV~1204 and Melotte~111~AV~1207.
This FOV was observed through a Johnson $V$ filter with an exposure time between 60  sec corresponding to a Nyquist frequency of 635 c/d.
After carefully analyzing the light curves and amplitude spectra of all the stars in this FOV we conclude that
none of the stars present significant variations attributed to intrinsic pulsations.

As was mentioned before the FOV~3 (Fig.~\ref{fig:field3}) was set in direction of Upgren 1.
We have observed four stars of this association namely 
HD~109509, HD~109530, HD~109542 and BD+37~2295 through a $y$-Str\"omgren filter
with a exposure time of 20 sec resulting in a Nyquist frequency of 2274 c/d.
After checking carefully the light curves and the amplitude spectra
we conclude that none of the stars is variable.  

Five stars in  FOV~4 (Fig.~\ref{fig:field4}) were checked for variability.
Three stars are presumable members of Melotte~111, namely
Melotte~111~AV~1224, Melotte~AV~1236, Melotte~111~AV~1248.
We found evident brightness changes on a time scale of few hours in Melotte~111~AV~1224. 
The adjacent stars in the field were found not to be variable stars. 
An in depth analysis of the light curve is given in the next section.

\begin{table*}
\caption{Log of observations.}
\begin{tabular}{lcccccr}

\hline

    UT Date 2009 &Fields & Start Time  & End Time &  Filter  & Exp. time &No. of \\
                  &    &(HJD 2454900+)&(HJD 2454900+)&     & (sec)&Images      \\
 \hline
       April 12 &FOV~1 (Fig.~\ref{fig:field1})   &     33.660060    &  33.956975   & $y$ & 30 & 1069 \\
         April 13 &FOV~2 (Fig.~\ref{fig:field2})  &      34.649227  & 34.974052    &  $V$  & 60&350 \\
        April 16 &FOV~3 (Fig.~\ref{fig:field3})   &     37.705917    &  37.989076 & $y$ &20 & 1195 \\
        April 17 &FOV~4 (Fig.~\ref{fig:field4})   &      38.631485     & 38.973245   & $V$&120 &230\\
        April 18 &FOV~4 (Fig.~\ref{fig:field4})   &   39.683163      &  39.683163  &   $V$&120&205 \\
        April 20&FOV~4 (Fig.~\ref{fig:field4})    &   41.661120     &   41.953617 &   $V$&120 &176  \\
        April 21 &FOV~4 (Fig.~\ref{fig:field4})    &     42.669725    &  42.669725  &   $V$ &120 &207\\
       
\hline
\end{tabular}
\label{tab:log}
\end{table*}

\section{Melotte~111~AV~1224 }
\label{sec:av1224}

\subsection{Light curve and period analysis}
\label{sec:per}
Due to variable nature of Melotte~111~AV~1224, we decided to monitor this star during all remaining nights.
A total number of 818 frames  were obtained through a $V$ Johnson filter. The exposure time was set to 120 sec.

\begin{figure}
\includegraphics[width=8.0cm,height=10.0cm]{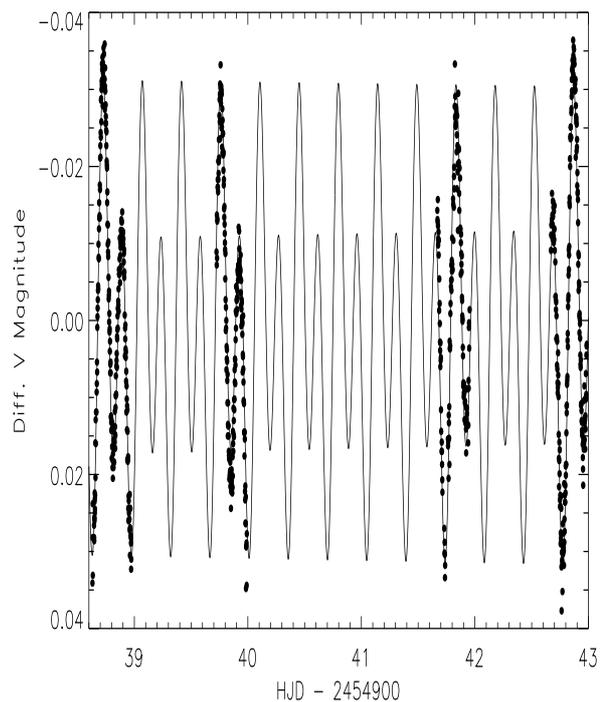} 
\caption{Timed light curve of Melotte~111~AV~1224. The continuous line is a synthetic light
curve generated with the main and harmonic frequencies.} \label{fig:curve}
\end{figure}

The light curve of Melotte~111~AV~1224 in the time space is shown in Fig.~\ref{fig:curve}.
As can be seen the light curve is not sinusoidal, but is strictly periodic.
The period of the light curve was derived by using the Period04 program \citep{lenz}.
The frequency spectrum reveals two peaks, $f_{1} \sim 2.9$~d$^{-1}$ and
$2f_{1} \sim 5.8$~d$^{-1}$.  The derived period is $P= 0.345895 \pm 0.000024$ days.

The following ephemeris was derived from the computed period of the binary system:

\begin{equation}
HJD_{\rm max}\,\,\,I = 2454938.728637  + 0.345895 \times E
\end{equation}


\noindent where the reference epoch was chosen to be the initial HJD time of the light curve.

Apart from calculating the light elements using the period derived with the Period04 program, 
we have also explored other means by which to determine the period of Melotte~111~AV~1224.
To do so, we determined individual time of maximum from the photometry data via the method of \cite{kwee}.
A total of 6 timings were obtained (4 primary eclipses and 2 secondary eclipses) which 
are listed in Table~\ref{tab:max}. These were then used to establish a period and a reference epoch by
solving for a linear ephemeris using standard least-squares techniques. 
Primary and secondary maxima were adjusted simultaneously and the orbit was assumed to be circular.
The resulting period and epoch are given by $P=0.3454  $ and $T_{0}=2454938.7244$ respectively.
The period is in agreement with that derived with the Period04 program. However as the related uncertainties
 are larger,  we adopt the ephemeris (1) for the remainder of the paper.

\begin{table*}[!t]\centering
  \setlength{\tabcolsep}{1.0\tabcolsep}
 \caption{Times of maxima for Melotte~111~AV~1224.}
  \begin{tabular}{lccr}
\hline
HJD      &       Type       &   E     &   O-C       \\
       
\hline
 $2454938.728637  \pm 0.000168$  & I & 0.0 & 0.000000 \\
$2454938.887158  \pm 0.000297$& II&   0.5 & -0.014426 \\
  $2454939.761764  \pm 0.002490$& I  & 3.0 &-0.004557\\
$2454939.921642 \pm 0.000367$&II  &  3.5 &  -0.017626\\
   $2454941.839738 \pm 0.000670$&I &  9.0     &  -0.001950 \\
$2454942.869751 \pm 0.000209$&I &  12.0  & -0.009621 \\

\hline
\end{tabular}
\label{tab:max}
\end{table*}

\subsection{Str\"omgren photometry and spectroscopy}
\label{sec:strom}

In order to shed more light on the physical nature of  Melotte~111~AV~1224  
Str\"omgren photometry and low-resolution spectroscopy was performed. 

Str\"omgren photometric observations were secured in June 2010 by using
the 1.5-m  telescope with the six-channel Str\"omgren spectrophotometer.  
This photometer has been widely used recently for deriving 
physical parameters of open clusters  [e.g. \citet{pena1,pena2}]. 
The observing procedure is the same as explained in \cite{fox5}.
Briefly,  the observing routine consisted of five 10 s of
integration of the star from which five 10 s of integration of the
sky was subtracted. Along with Melotte~111~AV~1224  also observed were
some A-F type {\it Kepler} targets \citep{uytterhoeven}.
A set of standard stars was also observed to transform
instrumental observations into the standard system and to correct
for atmospheric extinction.  The instrumental magnitudes ($_{\rm
inst}$) and colours, once corrected from atmospheric extinction,
were transformed to the standard system ($_{\rm std}$) through the
well known transformation relations given by \cite{stromgren} :

\[ V_{\rm std} = A + y_{\rm inst} + B(b-y)_{\rm inst} \]

\[(b-y)_{\rm std} = C + D(b-y)_{\rm inst} \]

\[ m_{1,{\rm std}} = E + Fm_{1,{\rm inst}} + G(b-y)_{\rm inst} \]

\[ c_{1,{\rm std}} = H + Ic_{1,{\rm inst}} + J(b-y)_{\rm inst} \]

\[ H_{\beta,{\rm std}} = K + LH_{\beta,{\rm inst}} \]

\noindent where $V$ is the magnitude in the Johnson system, and the
$m_{1}$ and the $c_{1}$ indices are defined in the standard way:
$m_{1} \equiv (u-v) -(v-b)$ and $c_{1} \equiv (v-b) - (b-y)$.

The following indices in the Str\"omgren system for Melotte~111~AV~1224 were derived:
$V=13.395 \pm 0.018$, $(b-y)=0.473 \pm 001$,  $m_1=0.271 \pm 0.006$, and $c_1=0.292 \pm 0.001$,
$H_{\beta} = 2.575 \pm 0.083$.

\begin{figure}[!t]
\includegraphics[width=8cm]{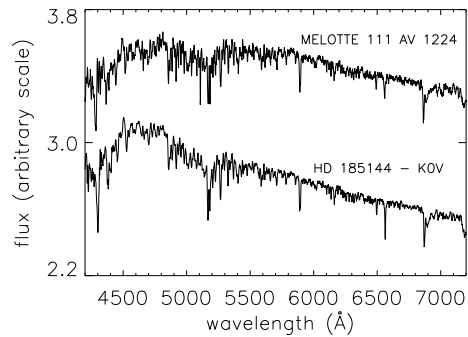} 
  \caption{Stellar spectrum of \astrobj{Melottte~111~AV~1224} and  HD~185144 a K0V star.}
  \label{fig:spectra}
\end{figure}

 Spectroscopic observations were conducted at the 2.12-m telescope in June 2011.
We used  the same equipment as explained by \cite{baran1}.
In particular, we used
Boller \& Chivens spectrograph installed in the Cas\-se\-grain focus of the telescope.
The 400 lines/mm grating with a blaze angle of 4.18$^{\circ}$ was used.
The grating angle was set to 7$^{\circ}$ to cover wavelength range from 4000\,{\AA} to 7500\,{\AA}.
A 2048$\times$2048 E2V CCD camera
was used in the observations.
The typical resolution of the recorded spectra is 8\,{\AA} and the dispersion amounts to
1.8\,{\AA} per pixel. The reduction procedure was performed with the standard routines
of the IRAF package. 
The spectral type was derived by comparing the
normalized spectrum of Melotte~111~AV~1224 with those of well classified stars taken on the same night. 
We have assigned a spectral type of K0V to
\astrobj{Melotte~111~AV~1224}. 
Fig.~\ref{fig:spectra} shows the reduced spectrum of Melotte~111~AV~1224 and the
spectrum of a standard star of spectral type K0V for comparison.

\subsection{Physical parameters from Str\"omgren photometry and spectroscopy}
\label{sec:par}

We have used the standard $uvby-\beta$ indices
 to estimate the unreddened colours of Melotte~AV~1224. The
UVBYBETA\footnote{Written by T.T. Moon at University London in 1985 and based
on calibrations of \citealt{moon}} IDL code was implemented to derive the following intrinsic colours:
$E(b-y)=0.022$,$(b-y)_{0}=0.451$ $m_{0}=0.278$,$c_{0}=0.288$,
$M_{V}=5.35$ which lead to  $T_{\rm eff} =5450 \pm 300$  K and $R=0.92 \pm 0.2$ $R_{\odot}$. 
A similar value of $T_{\rm eff}$ is obtained using the calibrations of \cite{ramirez}
for FGK stars.
The intrinsic colours 
are consistent with an early K-type star in agreement with our spectroscopic
classification. As a reference, the atmospheric parameters
of the standard star HD~185144  are the followings:
$T_{\rm eff} = 5318$ K, $\log\, g = 4.59$ cm/s$^{2}$ \citep{takeda}

The distance moduli of Melotte~111~AV~1224 amounts to 8.045 mag which
leads to a distance of 392 pc implying  that
Melotte~111~AV~1224 is  rather located behind the Molotte~111 open cluster. 

\subsection{Light curve modeling}
\label{sec:mod}

\begin{figure}[!t]
\begin{center}
\includegraphics[width=8.0cm,height=10.0cm]{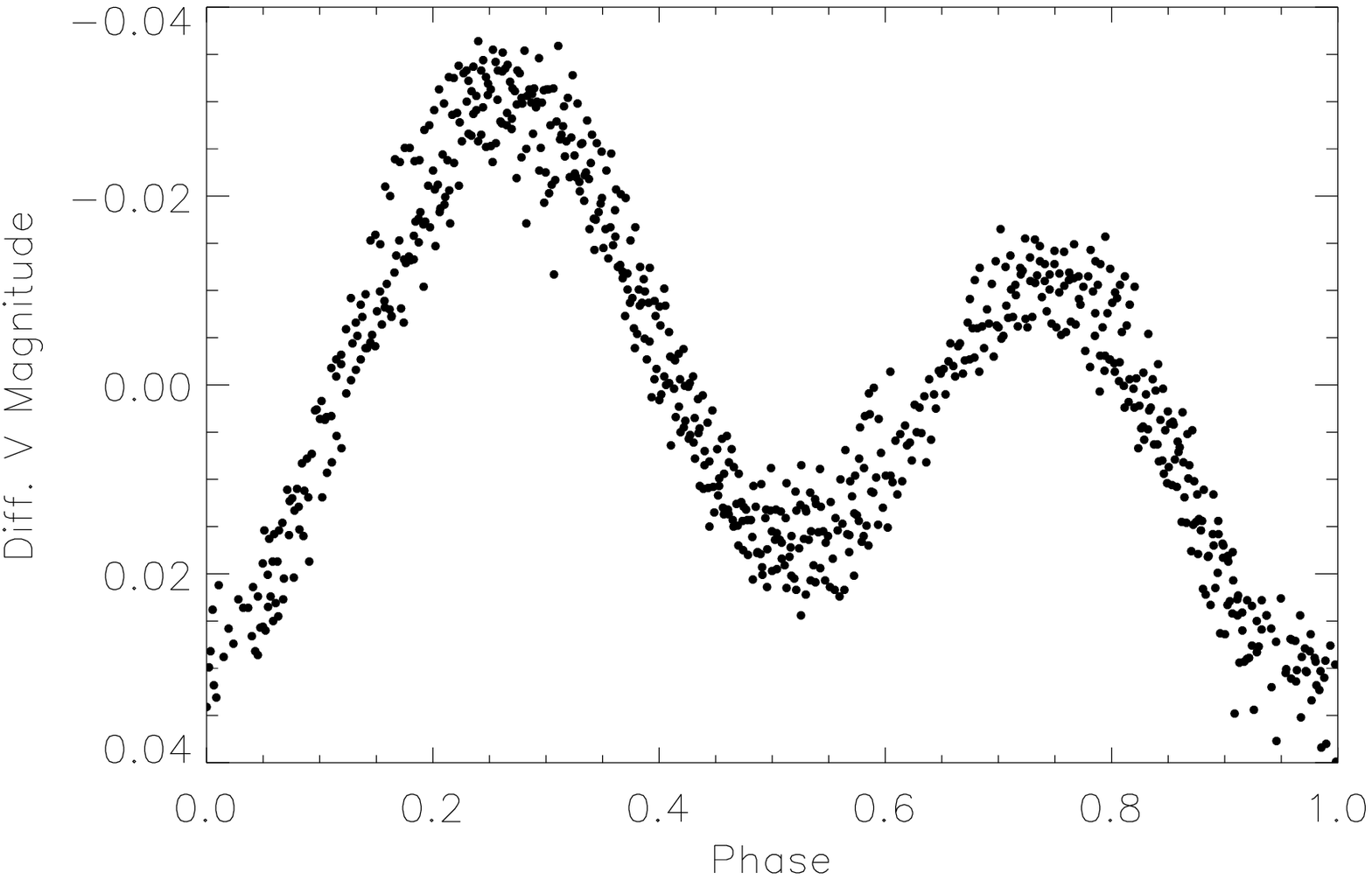}  
\caption{Phased light curve of Melotte~111~AV~1224.}
\label{fig:phase}
\end{center}
\end{figure}

Using the light elements derived in Sect.~\ref{sec:per}, we have constructed
the phased light curve of Melotte~111~AV~1224 shown in Fig.~\ref{fig:phase}. 
 In an effort to gain a better understanding of the binary system and determine its physical
properties, we have analyzed the phased light curve 
with the software PHOEBE V.0.31a (PHysics Of Eclipsing BinariEs,
\citealt{prsa}). PHOEBE is a software package for modeling
eclipsing binary stars based on the Wilson-Devinney code \citep{wilson}.
It permits creation of a synthetic light curve that would best fit
the observational data by  adjusting interactively the orbital and stellar 
parameters through a user interface friendly.
We averaged the phases and magnitudes every three points and use this binned light curve as input in 
the PHOEBE code.

In the light curve analysis we assume a few fixed parameters during
the fitting process: the temperature of the primary star, based on 
the Str\"omgren photometry (Sect.~\ref{sec:par}) is set to $T_{1}=5450$ K;
the period of the system and zero time $HJD_{0}$ were obtained from equation (1).
The logarithmic limb-darkening coefficients and bolometric limb-darkening coefficients,
 were determined from tables by \cite{vanhamme} for the primary and secondary 
components, respectively. Standard values of bolometric albedos \citep{rucinski}, and the gravity-darkening 
coefficients \citep{lucy} for radiative and convective envelopes were used.
The adjustable parameters in the light
curves fitting were the orbital inclination $i$, the surface 
potentials $\Omega_{1}$ and $\Omega_{2}$ , the effective temperature
of the secondary $T_{2}$, the mass ratio $q$ and luminosity of the primary component.

\begin{figure}[!t]
\begin{center}
\includegraphics[width=8.0cm,height=10.0cm]{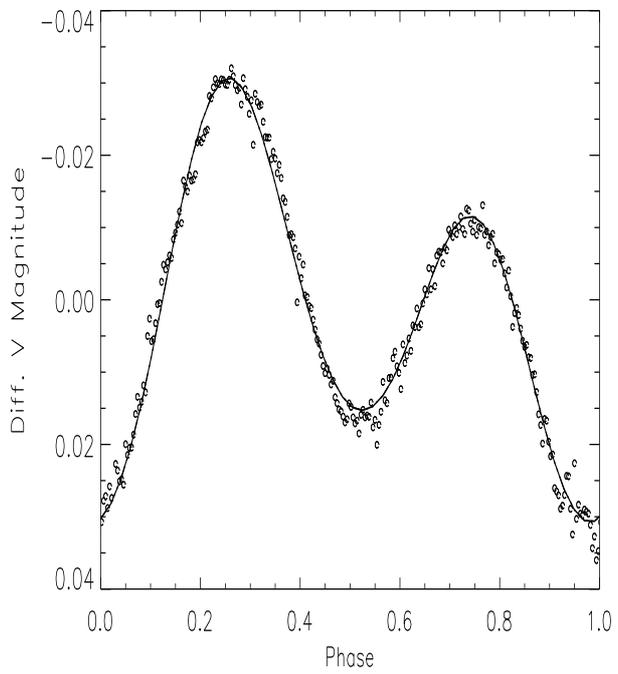}
\caption{The binned observational light curve in $V$  band (circles) and the best theoretical fit model (solid line)   
 of Melotte~111~AV~1224.}
\label{fig:phasebin}
\end{center}
\end{figure}

The shape of the light curve of Melotte~111~AV~1224 resembles those of both systems classical Beta Lyrae (EB)
and W-UMa, with a difference in amplitude of the primary and secondary eclipses and
 with no clear beginning and end of the eclipses. 
The difference in eclipse amplitudes strongly suggests significant
deformation of the components and perhaps some degree of contact.  Consequently, we performed fits in
overcontact mode (Mode-3), in semidetached mode  with the primary filling its Roche lobe (Mode-6) and
and in semidetached mode with the secondary component filling its Roche lobe (Mode-7).
As the primary and secondary eclipses occur at almost 0.5 phase interval suggesting
a  circular orbit,  we have set the eccentricity $e =0.0$.
During the fitting process the iterations were carried out automatically until convergence, 
and a solution was defined as the set of parameters for which the differential corrections
were smaller than the probable errors and the smallest $\chi^{2}$ was obtained.
With all configurations two spots on the secondary component have been considered 
during the fitting process to account for a slight asymmetry in the light curve.
Our analysis reveals that astrophysically reasonable solutions are obtained with either configuration
overcontact  or semidetached  with the secondary filling its Roche lobe,
even though the smallest $\chi^{2}$ is achieved in overcontact configuration.
Figure~\ref{fig:phasebin} depicts our best-fit theoretical light curve (solid line)
fitted to the observational data (circles) in overcontact  configuration mode.
The full list of fitted, absolute and spots parameters is given in Table~\ref{tab:par}.
For comparison in Table~\ref{tab:par1}  the fitted parameters in semidetached configuration are listed. 
The uncertainties assigned to the adjusted parameters are the 
internal errors provided directly by the code. 

Briefly, our results show that the Melotte~111~AV~1224 system appears to have
a mass ratio of $q \approx 0.21$, and low inclination angle of $i \approx 45^{\circ}$, and
a secondary star temperature of $T_{2} \approx 4200 $ K.  A temperature difference
of $\sim$ 1000 K between the two components is obtained which is consistent with the 
different eclipse depths observed in the light curves.

\begin{table*}[!t]\centering
  \setlength{\tabcolsep}{1.0\tabcolsep}
 \caption{Solution parameters for Melotte~111~AV~1224 in overcontact configuration.} \label{tab:par}
  \begin{tabular}{llrcc}
\hline
\hline
Parameters & &  &System &           \\
 &&Primary component& &Secondary Component\\
\hline
& &&& \\
 $HJD_{0}$ [days] & & &38.631482  &\\    
Orbital period -- $P$ [days]&&& 0.345894 & \\
Semi-major axis $a$ [$R_{\odot}$]  &       &  & $3.44 \pm 0.46$ & \\
Mass ratio $q=m_{2}/m_{1}$& & & $0.21 \pm 0.05$ & \\
Binary orbit inclination $i$ [$^{\circ}]$& & &$45 \pm 1$ & \\
Binary orbit eccentricity $e$& & & 0.0& \\
&&&&\\
Effective temperature [K]  &  &$T_{1}=5450.0$&& $4215 \pm 112$  \\
Surface potential  & &$\Omega_{1} = 2.414 \pm 0.005 $&& $\Omega_{2}=2.445 \pm 0.002$ \\
Bolometric albedo   & &$A_{1}=0.60$&&$A_{2}=0.60$\\
Exponent in gravity brightening &&$g_{1} = 0.32$&&$g_{2}=0.32$\\
Bolometric limb darkening coefficient &&$x_{1}=0.5$&&$x_{2}=0.5$\\
Linear limb darkening coefficient& &$y_{1}=0.5$&&$y_{2}=0.5$\\ 
&&&&\\
 &&&Absolute parameters&\\ 
Potential of Lagrangian point  & &$\Omega(L_{1})=2.261$&&$\Omega(L_{2})=2.126$\\
Mass [$M_{\odot}$]                      &    &$M_{1}=0.968$&&$M_{2}=0.204$\\
Radius [$R_{\odot}$]                    & &$R_{1}=1.037$&&$R_{2}=0.462$\\
Bolometric absolute magnitude & & $M_{1}^{\rm bol} = 6.677$& & $M_{2}^{\rm bol} = 4.961$ \\
Surface gravity $\log\,g$              & & $Log (g_{1}) = 4.392$&& $Log (g_{1}) = 4.420$\\
Luminosity [$L_{\odot}$]              & &1.622& &1.659\\
                             & &      & Spot parameters  &      \\
\hline
       &Colatitude &   Longitude  &   Radius    & Temperature factor\\
       &  [$^{\circ}$]&[$^{\circ}$]&[$^{\circ}$]& ${T_{\rm spot}}/{T_{\rm surf}}$\\
\hline
Spot 1  &18  &   300  &     44& 0.7\\
Spot 2  &15  & 180 & 50  &  0.7 \\
\hline

\end{tabular}

\end{table*}

\begin{table*}[!t]\centering
  \setlength{\tabcolsep}{1.0\tabcolsep}
 \caption{Solution parameters for Melotte~111~AV~1224 in semidetached configuration.} \label{tab:par1}
  \begin{tabular}{llrcc}
\hline
\hline
Parameters & &  &System &           \\
 &&Primary component& &Secondary Component\\
\hline
& &&& \\
 $HJD_{0}$ [days] & & &38.631482  &\\    
Orbital period -- $P$ [days]&&& 0.345894 & \\
Semi-major axis $a$ [$R_{\odot}$]  &       &  & $2.31 \pm 0.49$ & \\
Mass ratio $q=m_{2}/m_{1}$& & & $0.19 \pm 0.06$ & \\
Binary orbit inclination $i$ [$^{\circ}]$& & &$40 \pm 1$ & \\
Binary orbit eccentricity $e$& & & 0.0& \\
&&&&\\
Effective temperature [K]  &  &$T_{1}=5450.0$&& $4005 \pm 100$  \\
Surface potential  & &$\Omega_{1} = 2.349 \pm 0.005 $&& $\Omega_{2}=3.622 \pm 0.06$ \\
Bolometric albedo   & &$A_{1}=0.60$&&$A_{2}=0.60$\\
Exponent in gravity brightening &&$g_{1} = 0.32$&&$g_{2}=0.32$\\
Bolometric limb darkening coefficient &&$x_{1}=0.5$&&$x_{2}=0.5$\\
Linear limb darkening coefficient& &$y_{1}=0.5$&&$y_{2}=0.5$\\ 
&&&&\\
 &&&Absolute parameters&\\ 
Potential of Lagrangian point  & &$\Omega(L_{1})=2.379$&&$\Omega(L_{2})=2.211$\\
Mass [$M_{\odot}$]                      &    &$M_{1}=1.064$&&$M_{2}=0.278$\\
Radius [$R_{\odot}$]                    & &$R_{1}=1.129$&&$R_{2}=0.617$\\
Bolometric absolute magnitude & & $M_{1}^{\rm bol} = 7.656$& & $M_{2}^{\rm bol} = 4.778$ \\
Surface gravity, $\log\,g$             & & $Log (g_{1}) = 4.359$&& $Log (g_{1}) = 4.301$\\
Luminosity [$L_{\odot}$]                & &1.622& &1.659\\
                             & &      & Spot parameters  &      \\
\hline
       &Colatitude &   Longitude  &   Radius    & Temperature factor\\
 &  [$^{\circ}$]&[$^{\circ}$]&[$^{\circ}$]& ${T_{\rm spot}}/{T_{\rm surf}}$\\
\hline
Spot 1  &6  &   100  &     20& 0.7\\
Spot 2  &4  & 110 & 35  &  2.0\\
\hline

\end{tabular}

\end{table*}

\section{Results and conclusions}
\label{sec:con}

A summary of a search for new short-period pulsating variables in direction of the
open clusters Melotte~111 and Upgren 1 has been presented.
35 stars were checked for variability in four observed fields.
We  did not confirm the variability in the stars
NSV~5612 and NSV~5615  considered as variable stars of
the Melotte~111 open cluster. On the contrary, the star  Melotte~111~AV~1224 was found
to be a new eclipsing binary star. 
Follow-up CCD observations  of  Melotte~111~AV~1224
allowed us to estimate the orbital period and ephemeris of the system.
Based on Str\"omgren standard photometry  and low-resolution spectra we conclude
that the primary component is most likely an early K-type dwarf. 
The analysis of the Str\"omgren standard photometry place it to 392 pc much more farther
that Melotte~111 open cluster ($\sim$ 87 pc).  Therefore,  Melotte~111~AV~1224 is not dynamically
associated with the Melotte~111 open cluster. This is consistent with the 
fact already pointed out in early investigations that the Melotte~111 open cluster 
has relatively few members and particularly that 
it is poor in low-mass stars.  
Although a classical Beta Lyrae (EB) binarity classification cannot be ruled out,
our analysis of the light curve of Melotte~111~AV~1224 revealed properties
similar in many respects to those of the W UMa systems, which are characterized
by having short orbital periods (0.2 - 0.8 d) and are composed of F-K type stars sharing a common envelope.
 However the evolutionary history
of the system is not clear due to the missing of radial velocity data. We have found
that both models, overcontact  and semidetached, systems fit the observed light curves
equally well. We think that the system is undergoing cyclic variations with
alternating phases of true contact and semidetached, but almost contact, phases.
During the contact phases the characteristic W-Uma light curve should be observed,
while during semidetached phases the surface temperature of the components should
be different, thus producing Beta Lyrae (EB) type light curve. 
Therefore we are probably seeing the semidetached phases of the system. 

To date our observations represent the most extensive work on Melotte~111~AV~1224.
Overall we believe that our results are the best that we can achieved based
solely on photometric observations made in the $V$ filter. For a better 
understanding of the properties, both spectroscopic observations and photometric
data at multiple wavelengths are needed.

\bigskip
{\bf \noindent Acknowledgments}

This work has received financial support from the UNAM via grant
IN114309.  Based on observations collected at the 0.84 m
telescope at the Observatorio Astron\'omico Nacional at San Pedro
M\'artir, Baja California, Mexico. Special thanks are given to the
technical staff and night assistants of the San Pedro M\'artir observatory.
We thank J. Miller for a careful proofreading of this manuscript.
 This research has made use of the SIMBAD
database operated at the CDS, Strasbourg (France).

\bigskip
{\noindent \bf References}
\medskip

\bibliographystyle{elsarticle-harv}

\end{document}